# Qiskit Quantum Circuits Posit Singlet state in Radical Pair-based Magnetoreception of Migratory Birds

Hillol Biswas[1,3,*]          Spyridon Talaganis[1,2,4]

[1] Department of Electrical and Computer Engineering, Democritus University of Thrace, Xanthi, Greece

[2] Physics Department, Lancaster University, Lancaster LA1 4YB, United Kingdom

[3] hillbisw@ee.duth.gr

[4] s.talaganis@lancaster.ac.uk

[*] Corresponding author.

## ABSTRACT

Quantum computing applications in diverse domains are emerging rapidly. Given the limitations of classical computing techniques, the peculiarity of quantum circuits, which can observe quantum phenomena such as superposition, entanglement, and quantum coherence, is remarkable. This capability enables them to achieve measurement sensitivities far beyond classical limits. Research on radical pair-based magnetoreception in migratory birds has been a focus area for quite some time. A quantum mechanics-based computing approach, thus unsurprisingly, identifies a scope of application. In this study, electron-nucleus spin quantum circuits for different geomagnetic fluxes have been simulated and run through IBM Qiskit quantum processing units with error mitigation techniques to observe the phenomenon. The results of different quantum states are consistent, suggesting singlet-triplet mechanisms that can be emulated, resembling the environment-enabling flights of migratory birds through generations of the avian species. The four-qubit model emulating electron-nucleus systems mimicking the environmental complexity outcome shows the sensitiveness to change of magnetic flux index, high probability of singlet-triplet dynamics, and upholding radical pair model states by the purity of the sub-system and full system outcome of coherence, the hallmark of singlet state dominance. The work involved performing fifty quantum circuits for different magnetic field values, each with one thousand and twenty-four shots for measurement, either in the simulator or on real quantum hardware, and for two error mitigation techniques, preceded by a noise model of a simulator run.

## I. INTRODUCTION:

The phrasal backstopping that nature is quantum is one of the reasons for advancing quantum computing research in the current computational regime. In his book "What is Life?", 1951, and lecture series, Schrödinger [1] made various inculcating and contemplative thoughts about the functional characteristics of DNA and possible lineage with quantum phenomenon. Many fields of physics and mathematics have found applications in biology, from the statistical techniques used in bioinformatics to the mechanical and factory-like qualities observed at the microscale within cells. This field is currently advancing at a rapid rate. Naturally, this development raises the question of whether quantum physics has any application in biology [2]. While the wondering debate on the classical and quantum world likely has existed for decades, if not more, the wonders of nature often leave us spellbound by the vivid dimensions that carry on since ages past across the globe. Avian migration is an incredible quagmire that has led research communities apart from ordinary public members to wonder and contemplate. Millions of migratory birds travel across continents, matching the earth's revolution that directly causes seasonal changes, and animals tend to adapt to this amazingly fascinating world of dynamics. Nonetheless, generations of this beautiful species carry on the tradition and legacy, a subtle phenomenon of nature. However, research communities have tended to address this from varied dimensions, and one of them is the background quantum phenomenon, which might be one of the driving forces behind it.

A flight path, Figure 1, of thousands of kilometers across the ocean, mountains, and ridges, and encountering possibly all of the diverse terrains, demands a time that is at least several weeks to months, depending upon the distance and places. This remarkable phenomenon, nevertheless, requires many variables of the computational regime, including time and direction, and ends up leaving the pattern of trend, seasonality, cyclicity, duration, flight map, natural phenomena such as weather and climate, and last but not least, sunlight and the Earth's geomagnetic field. Unsurprisingly, both light and geomagnetic fields are quantum phenomena!

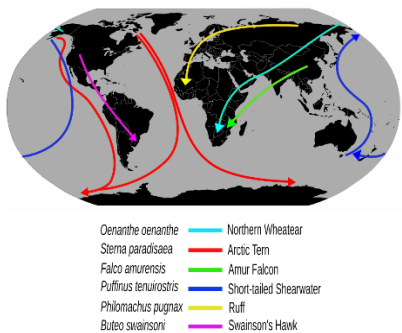

*Figure 1: Migratory Birds various Flight path*

Source: Wikipedia

It is now known that a wide range of animals, from birds to lobsters, utilize magnetic positional information for various reasons, including navigating towards specific objectives, adjusting food intake at strategic times during migration, staying within a suitable oceanic region, and maintaining course along migratory pathways [3].

The hypothesis of magnetoreception as a result of a chemical compass based on radical pair and process involving biogenic magnetic particles has been discussed [4],[5], [6]. Moreover, light-dependent magnetoreception for different spectrum [7] and its effects on migratory birds under different theoretical and experimental settings have also been studied. According to the Radical Pair model, magnetoreception is a process that depends on light. Migratory birds exhibit orientation in their migratory path when exposed to low monochromatic light from the short-wavelength portion of the visual spectrum. In the monochromatic illumination of higher intensity, they also had odd axial or other direction preferences [8].The Earth's magnetic field not only gives animals a source of directional, or "compass," information, but it also offers a potential source of positional, or "map," information that they could use to determine their location. The notion that animals use the Earth's magnetic field as a map has evolved from a controversial theory to a widely accepted principle of animal navigation in less than a generation. It is now known that a wide range of animals, from birds to lobsters, use magnetic positional information for several reasons, such as navigating towards particular objectives, modifying food intake at strategic points during a migration, staying within an appropriate oceanic region, and staying on course along migratory pathways [3].

## II. QUANTUM COMPUTING-BASED APPROACH TO MAGNETORECEPTION:

The emerging field of quantum biology explores how quantum mechanics influences biological processes. Animal quantum sensing is one of its most fascinating features; biological systems appear to utilize quantum processes to enhance sensory perception. This is especially important for vision, olfaction, and navigation. Animal Quantum Sensing Mechanisms is the area that takes stock of several animals that display sensory capacities that appear to function outside the realm of classical physics, possibly utilizing quantum mechanics, viz: Bird navigation, magnetoreception, olfaction.

The European robin is one of the many migrating birds that use the Earth's magnetic field for navigation. The Radical Pair Mechanism (RPM) is the leading theory, which stipulates that Birds' eyes include cryptochrome proteins excited by light, producing radical pairs. There is a quantum superposition of spin states, including these radical pairings. These states are influenced by the Earth's magnetic field, which gives direction information. The magnetic field may appear to birds as a visual overlay. Experiments on fruit flies, or Drosophila, confirm Cryptochrome's role in magnetoreception. According to quantum coherence studies, birds may retain entangled radical pairs for remarkably extended periods on the order of microseconds. In 2021, scientists created synthetic molecules with entangled electron spins to simulate magnetoreception.

Moreover, Quantum Tunnelling and Olfaction is another approach in which, according to Luca Turin's vibrational theory of smell, odor molecules can be recognized by quantum tunneling in addition to form. Thanks to chemical vibrations, electrons may tunnel across an olfactory receptor when a molecule binds to it. This mechanism might explain animals' ability to discriminate between similar molecules with distinct odours. Turin's team supported the quantum hypothesis by demonstrating that fruit flies can discriminate between isotopic variations (deuterated vs. normal molecules). Nowadays, electron tunnelling in olfactory receptors is simulated by computational models.

## Radical pair mechanism

Hore et al. provided an in-depth study entailing i) Mechanism of Radical Pairs (RPM) for magnetoreception. Photoexcitation creates radical pairs, which give rise to electron spin interactions that are influenced by magnetic fields. Reaction yields are influenced by the interconversion of the radical pair, which can exist in either singlet or triplet states and may function as a biological compass.

ii) Magnetoreceptors in Cryptochromes: Proteins known as cryptochromes are well-known for their role in light perception and regulating circadian cycles. They are excellent candidates for magnetoreception because they absorb light and produce radical pairs.

iii) Biophysical Aspects to Consider - Singlet-triplet transitions induced by hyperfine interactions in radicals enable sensitivity to weak magnetic fields. Effective magnetoreception depends on the lifetimes of the radical pairs and their capacity to maintain coherence.

iv) Anatomical and Biological Features—The retina is the most plausible site for magnetoreception due to its exposure to light. Cluster N, an active area of the brain during magnetic navigation, is where the neural processing of magnetic information occurs.

In summary, Magnetic fields are probably "seen" by birds as patterns. Studies have shown that magnetic field effects may be present in pure cryptochrome proteins. According to behavioral research, radio-frequency waves interfere with birds' magnetic compass, which depends on light. Electromagnetic noise and time-dependent magnetic fields can disrupt magnetoreception.

Since the 1970s, it has been established that applied magnetic fields can affect several chemical reactions. Radical pairs—transient radicals produced concurrently, with one electron spin on each radical—are the essential species. They have two unique characteristics: the electron spins remain far enough from thermal equilibrium for an extended period, making the kBT objection moot, and their chemical destiny is determined mainly by weak (kBT) magnetic interactions through their spin correlation. A remarkable amount of information on the magnetic characteristics, kinetics, and dynamics of radicals and their reactions has been provided by experimental and theoretical investigations of the "radical pair mechanism" over the past 30 years, as reviewed in previous studies. The term "spin chemistry" has been used to describe this branch of study. Key features of magnetoreceptors are, among other things, hyperfine interactions and cryptochromes act as magnetoreceptors [9].

### Avian magnetoreception

Animals that sense the Earth's magnetic field use it to obtain navigational information. The vector shows directions, and the magnetic inclination and intensity decrease from the magnetic poles to the magnetic equator. Magnetic declination may also be used as part of the navigational map. Birds use the geomagnetic field in two ways: the vector serves as a compass, and additional parameters—likely magnetic intensity—appear crucial to the navigational "map" required for long-distance travel. Although many features of the receptive processes have been detailed in recent years, it is still unclear how birds feel these characteristics [10]. Magnetoreception, the capacity of certain migrating animals to use the Earth's magnetic field for navigation, is known as magnetoreception. The specific mechanism and its characteristics appear to differ significantly among animals [2], though migratory birds are believed to have used this for generations. The simultaneous production of RPs from non-radical precursors, sometimes called geminate RPs, is arguably the most popular pathway to spin correlation. Geminate radical ion pairs ([D•+/A•−]) are created through electron transfer processes between neutral, diamagnetic donors, D, and acceptors, A. Many biological examples can be triggered by light [11].

Animals may navigate and orient themselves using the directional and positional information the Earth's magnetic field provides. There is increasing evidence that night-migratory songbirds may use geomagnetic information to estimate their approximate location on Earth, and magnetic information may be a valuable component of a map sense, particularly over longer distances as numerous studies indicate. Therefore, it is well-documented how night-migratory birds behave in response to geomagnetic cues, and many of them appear to possess both a magnetic map and a magnetic compass. On the other hand, one of the most important unresolved issues in sensory biology is still comprehending the underlying biophysical mechanisms [12].

What kind of radical pair might be involved in this mechanism is currently unknown. The main suspect is a sequence of radical-pair reactions known to occur in cryptochromes, which may generate a visual signal that the host species uses for navigation, as they are found in the eye. To demonstrate that the radical-pair reaction products are sensitive to the inclination of the external field and can replicate the disruptive effect of time-dependent external magnetic fields at radio frequencies, basic models of these types of radical pairs using highly anisotropic nuclear spin configurations are adequate [2]. A molecule with an odd number of electrons is called a radical. Two radicals produced simultaneously, typically through a chemical reaction, make up a radical pair. Take methane (CH4), a molecule with a carbon atom bound to four hydrogen atoms in a tetrahedral configuration. The carbon contributes six of its 10 electrons, whereas the hydrogens each contribute one. Eight electrons are involved in forming carbon-hydrogen bonds, with two electrons per bond, whereas two electrons surround the carbon nucleus. A radical pair made up of a hydrogen radical, H·, referred to as a hydrogen atom, and a methyl radical, · CH 3, is produced when one of the bonds is broken, leaving both ensuing fragments uncharged. The dots show one odd electron per radical. Because the electron, like the proton and neutron, possesses a property called spin, or spin angular momentum, radicals are magnetic. It is easy to picture the electron as a tiny, spherical entity that rotates on its axis, resembling a little planet. Like an electrical current in a wire loop, one could think that the charged and moving electron would produce a magnetic field. However, quantum objects do not behave conventionally, and spin is a property of quantum mechanics. Like mass or charge, spin is best understood as a property that some particles possess while others do not [12]. Spectroscopic observation of a carotenoid–porphyrin–fullerene model system to measure the anisotropic chemical response necessary for its function as a chemical compass sensor and to show that the application of #50 µT magnetic fields changes the lifetime of a photochemically formed radical pair. These tests demonstrate the viability of chemical magnetoreception, which also sheds light on the dynamic and structural design elements necessary for the best possible detection of the Earth's magnetic field direction [13].

Although numerous fruitful studies have been conducted over the past 20 years, many unanswered questions and conflicting results remain to be addressed. The fundamental procedures for identifying directions, if they adhere to the radical pair model, seem to be relatively well understood; nevertheless, it is still unclear how and where this data is sent and processed in the end [10]. The formation flights of migratory birds inspired the quantum-based avian navigation optimizer algorithm (QANA), a novel differential evolution (DE) method proposed in a study aimed at developing a unique algorithm. It is motivated by the remarkable precision of migratory birds' navigation on long-distance aerial itineraries. To efficiently traverse the search space using the suggested self-adaptive quantum orientation and quantum-based navigation, which consists of two mutation techniques, DE/quantum/I and DE/quantum/II, the population is divided into many flocks in the QANA. Every flock, except the first iteration, is allocated to one of the quantum mutation techniques using a recently proposed success-based population distribution (SPD) approach. In the meantime, a novel communication topology called V-echelon is used to distribute the information flow among the population [14]. A paper has estimated the average lifespan of the radical pair-taking component in bird magnetoreception using the findings of two distinct behavioral studies conducted with European Robins. Our estimate of the lifetime is a few microseconds, which is in good agreement with experiments, in contrast to a recent study that estimated the average lifetime to be close to 100 µs based on the results of only one behavioral test. The most significant outcome of this work is the identification of a parameter regime in which the presence of an environment improves the chemical compass's performance [15].

It is still primarily unknown what transduction mechanisms enable specialized sense cells to transform changes in magnetic fields into electric signals. Birds have been a preferred model among species that have been demonstrated to sense Earth-strength magnetic fields because behavioural tests indicate that magnetic fields significantly impact their ability to identify their direction. Numerous species, including bacteria and vertebrates, contain the ferromagnetic mineral magnetite. The trigeminal nerve in birds has been linked to both single-domain magnetite and superparamagnetic (SPM) magnetite[16].

Moreover, Obstacles and Unanswered Questions remain as it is still unclear the precise biochemical routes connect radical pairs to brain impulses. More research is required to determine how cryptochrome-binding partners stabilize radical pairs. Further research is needed to understand spin relaxation processes and how they affect the compass mechanism's resilience [12].

This paper contributes to design and implement an electron-nucleus spin-based iterative quantum circuits in IBM Qiskit Simulator/QPU for magnetoreception, and to interpret its physical significance. A Qiskit-based approach to model magnetoreception for observing quantum phenomenon of entanglement, superposition and coherence is the paper's novelty.

### Framing of Radical Pair-based Magnetoreception Quantum Circuit:

Theoretical studies on radical-pair-based magnetoreception focus on three main aspects: entanglement, compass precision, and realistic radicals [12]. Entanglement is supposedly a key concept in the Role of Quantum Entanglement in the Radical Pair Mechanism.

Quantum computing, inspired by phenomena in nature, delves into an interesting domain; it started with Richard Feynman's [17] pondering and vision for the simulation of physics, recognizing the limitation of computation by the Turing machine. Quantum Turing machine [18], quantum circuit [19], quantum computation [20]Quantum parallelism and the research endeavour is just getting bigger and bigger [21], [22], [23]. In subsequent years, the Deutsch algorithm [24], [25] for the 2-qubit system, the Deutsch-Jozsa algorithm, the Bernstein-Vazirani algorithm, the Grover and Shor algorithms, and others, such as endeavours in computational complexity, search algorithms, and encryption, have rapidly developed. Quantum simulators viz. QCS [26], potential quantum register base [26], the recent advent of Qiskit by IBM provides leverage to explore the unknown research area and scope. It also branches out into quantum machine learning in the follow-up.

Earth's Magnetic Field (Typical Ranges) vary from location to location and at heights, presumably. At the equator, it is circa 25 µT (0.000025 T). At mid-latitudes, the magnetic field strength ranges from 45 µT (0.000045 T). At the poles, the magnetic field strength is approximately 65 µT (0.000065 T).

We will set the B-field values for a range from 10 to 100 µT for this study.

Spin entanglement and magnetic field interactions are essential to the radical pair mechanism. A dephasing channel is used to introduce decoherence, which simulates interaction with the surroundings. To determine how long the system maintains coherence, the aim is to measure the end state. To determine the duration of entanglement, the aim is to measure the decoherence times. Concept of a Quantum Circuit for Olfaction using Quantum Tunnelling: A quantum mechanical mechanism called "electron tunnelling" allows electrons to flow past a potential barrier. According to the olfaction theory, tunnelling rates are influenced by the frequencies at which various molecules vibrate. The quantum circuit should use qubits to represent electron states. Further, the tunnelling probability can be modified by simulating molecular vibrations using quantum gates. Then, calculate the likelihood that tunnelling will be successful by the quantum circuit. Building Quantum Circuits and Implementation in IBM QPU using Qiskit, [27] By demonstrating that modest decoherence can be advantageous, it defies accepted wisdom and offers a novel perspective on quantum effects in biological systems [15]. However, in the following two phenomena, only RPM are being considered for quantum circuit simulation.

The radical pairs originate in pure singlet or triplet spin states, when formed by chemical processes from precursors of the respective multiplicity, or else, when formed by free radical encounters, will assume this type of spin polarization due to multiplicity selection in a recombination process. In general, the pure spin states are not eigenfunctions of the Hamiltonian $H_{st}$ and will consequently undergo coherent evolution in time. This process is termed "spin evolution" [28]. This review discusses singlet and triplet spin states in radical pair mechanisms—important for magnetoreception and spin dynamics.

$$|S\rangle = \frac{1}{\sqrt{2}}(|\uparrow\downarrow\rangle - |\downarrow\uparrow\rangle) \qquad (1)$$

$$|T_+\rangle = |\uparrow\uparrow\rangle \qquad (2)$$

$$|T_0\rangle = \frac{1}{\sqrt{2}}(|\uparrow\downarrow\rangle + |\downarrow\uparrow\rangle) \qquad (3)$$

$$|T_-\rangle = |\downarrow\downarrow\rangle \qquad (4)$$

Considering an approach, a quantum circuit comprising two electrons and nuclear for capturing hyperfine coupling and coherence in the full system is built. Approach comprises 4 qubits 2 each for the electron and the nucleus for simulating a radical pair mechanism (RPM) in Qiskit. While a 3-qubit circuit could resemble 2 electrons with 1 nucleus, however, considering coherence due to environmental effects a rather complex circuit has been envisaged. Considering, earth magnetic field, 50 different Earth geomagnetic field strengths (in Tesla) from 10 µT to 100 µT have been put in a data frame with value of Bz and derived variable Θ as below:

Physical constants, gyromagnetic ratio = $1.76*10^{11}$ rad/s/T for 1 µs interaction time, t = 1e-6, converted Bz to rotation angle θ (radians), Larmor precession angle, spin in the external magnetic field is given by

$$\theta = \gamma_e * B_z * t \qquad (5)$$

*Table 1: Tabulation of Magnetic flux and Larmor precession angle*

|    | Bz       | theta     |
|----|----------|-----------|
| 7  | 0.000023 | 4.021239  |
| 21 | 0.000049 | 8.545132  |
| 49 | 0.000100 | 17.592919 |
| 28 | 0.000061 | 10.807079 |
| 18 | 0.000043 | 7.575726  |

The quantum circuit comprises of singlet, $\psi^-$ (singlet state $|S\rangle$) from Bell states with Hadamard Gate (H) which puts the states in superposition and CNOT gate (cx) in Qiskit,

$$H = \frac{1}{\sqrt{2}}(|01\rangle - |10\rangle) \qquad (6)$$

The x gate flips the state of qubit from 0 to 1 for qubit 1 as state preparation. The hyperfine interaction between nuclear and electron coupling is considered in qubits 2 and 3 with the CNOT gate,

$$CNOT = \begin{vmatrix} 1 & 0 & 0 & 0 \\ 0 & 1 & 0 & 0 \\ 0 & 0 & 0 & 1 \\ 0 & 0 & 1 & 0 \end{vmatrix} \qquad (7)$$

Pauli Z gate is applied for a phase flip. Entanglement interaction from the magnetic field with control qubit as nuclear and target qubit as electron was introduced with RZ gate as time evolution. If $H = \frac{\omega}{2}Z$

$$U(t) = e^{iHt} = e^{i\frac{\omega t}{2}Z} \qquad (8)$$

The Z rotation $U(t) = RZ(\theta) where \theta = \omega t$

The circuit ends with classical measurements of the qubits. Quantum data encoding is of different types viz. amplitude encoding, phase encoding, state encoding, and angle encoding, that can be used for different purposes depending on the criteria. Using phase encoding, the magnetic field values are input in the quantum circuits [29]. Figure 2 depicts the first quantum circuit with corresponding values from the data frame whereas Figure 3 reveals the same quantum circuit for different input of the data frame corresponding to second row. There are the first two quantum circuits of the total fifty number of circuits build based on data frame magnetic field values.

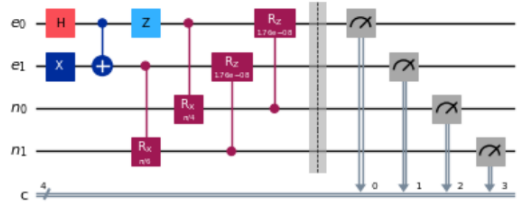

*Figure 2: 1st quantum circuit comprising 2 electron and 2 nuclei*

Similarly, the 2nd quantum circuit with θ = 2.08 is given in Figure

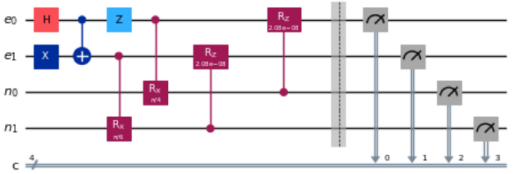

*Figure 3: 2nd quantum circuit comprising 2 electron and 2 nucleus*

### III. RESULTS

The multi-vector plot, Figure 4, comprises of Bloch sphere of all four qubits. The qubit 0 and 1 i.e., the electron spins are in mix state with no visibility of the red arrow. Its presence implies that entanglement between electrons still exists, and that the nuclear system is indicatively entangled with the electrons. If the system is mixed, the Bloch sphere representation does not show a well-defined state vector (arrow). Instead, the visualization may appear empty, as the Bloch sphere primarily represents pure states.

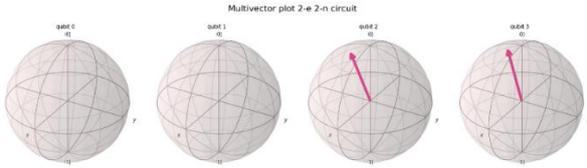

*Figure 4: Multivector plot of the quantum state*

The array to latex matrix representation yields as $-0.6532814824|0001\rangle+0.6830127019|0010\rangle+(-0.0377600947+0.2679505177i)|0101\rangle+(0.0255381624-0.1812221049i)|1010\rangle$.

State $|0001\rangle$ is down spin second nucleus amplitude -0.652814824, state $|0010\rangle$ is down spin first nucleus amplitude 0.6830127019, state $|0101\rangle$ is second spin down electron and nucleus amplitude $(-0.0377600947+0.2679505177i)$ and state $|1010\rangle$ is first electron and nucleus spin down amplitude $(0.0255381624-0.1812221049i)$. The total probability is 1 by summing up the corresponding amplitudes of states.

Purity of electron subsystem yields as 0.898 i.e., less than 1 indicating mix state. However, Purity of all four qubits i.e., electron-nucleus systems full yields 1 as pure state. It implies the whole system is pure however, the electron subsystem is in mixed state ratifying the multi-vector and other plots. Electron spins are entangled with nuclear spins. This drives the singlet-triplet oscillations.

The state city plot and the Hinton plot, Figure 5 & Figure 6 reveals the density matrix comprising of real and imaginary parts. Significant off-diagonal bars indicate non-trivial correlations likely resembling entanglement due to spin-spin and hyperfine couplings. The imaginary present parts indicate coherence among the qubits.

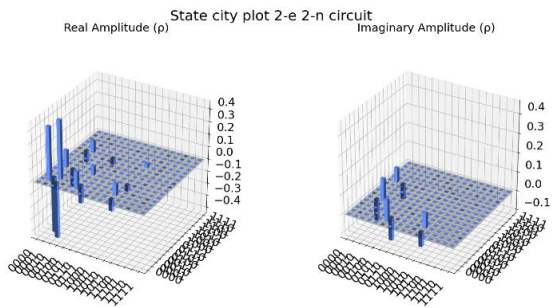

*Figure 5: State city plot of real and imaginary parts*

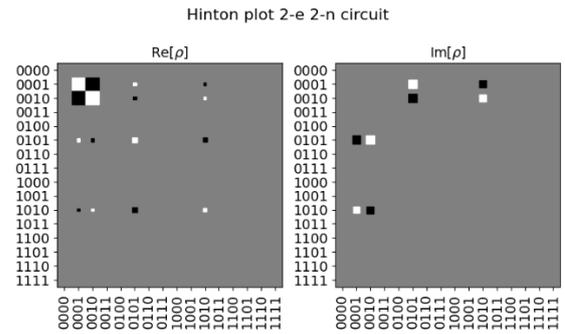

*Figure 6:Hinton plot of real and imaginary part of the quantum states*

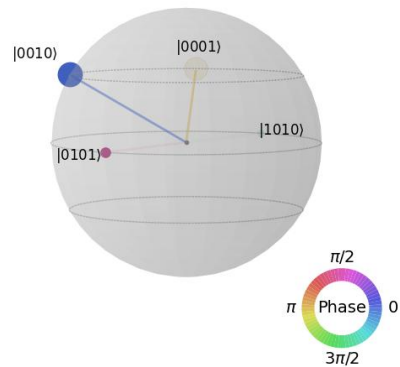

*Figure 7:Qsphere of the quantum state of all the qubits*

The qsphere, Figure 7, depicts the quantum states of multi-qubit systems with the use of point in the sphere. The dominant basis states are visible using Dirac notation. The dark blue colour represents the dominant state amplitude and the overall quantum coherence of the full electron-nucleus system comprising of two sub-systems. Figure 8, the coefficients indicate a radical pair undergoing coherent evolution in the presence of an external magnetic field, causing singlet-triplet mixing of the system.

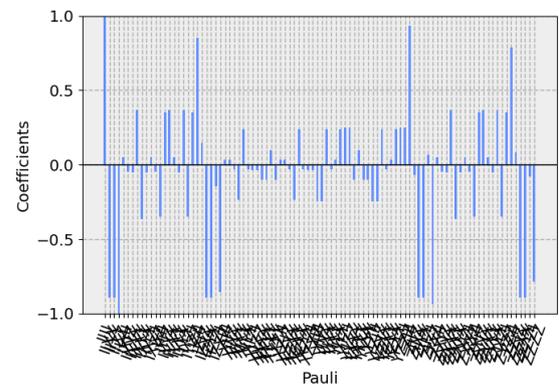

*Figure 8: Paulivectors coefficients*

Table 2 depicts the comparison of the built quantum circuit and the transpiled quantum circuit for customizing to hardware configurations. Introduction of mapped gates are specific to the IBM quantum hardware requirement before running on quantum processor units.

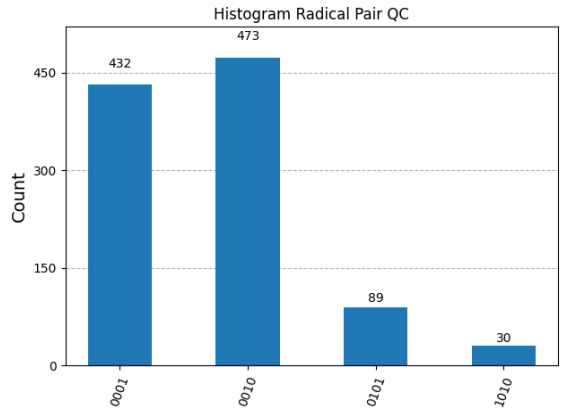

*Figure 9: AerSimulator results histogram of states*

Figure 9 reveals a typical circuit count plot, which is iterated for all the magnetic field values of the pandas data-frame, as in Table I. The 0001 and 0010 dominant values indicate that the electrons are entangled i.e., singlet state for the very circuit coupled with nuclear spins which the total spin states evolve with time. Figure 10 reveals the distribution plot for all the circuits after measurements with Aer Simulator with 1024 shots. It is remarkable to note the consistency with the single quantum circuit simulator run count plot, Figure 9, with Figure 10 for multiple runs for all 50 quantum circuits.

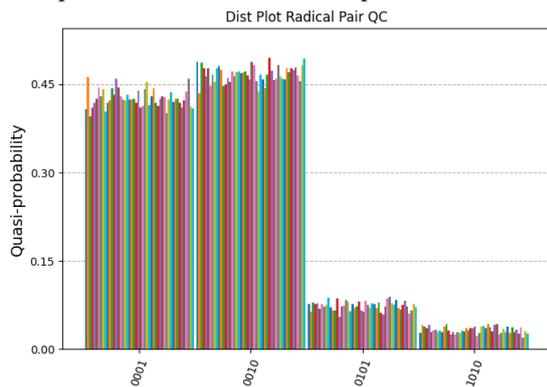

*Figure 10: Distribution plot of the aer simulator of all the circuits*

This plot, Figure 11 shows how the singlet yield varies as a function of the magnetic field strength, or equivalently, the rotation angle θ (which is proportional to Bz) in the quantum simulation of the radical pair mechanism. The magnetic field index is plotted against the singlet yield, which is the probability of the system ending in a singlet state after the measurement by Qiskit simulator.

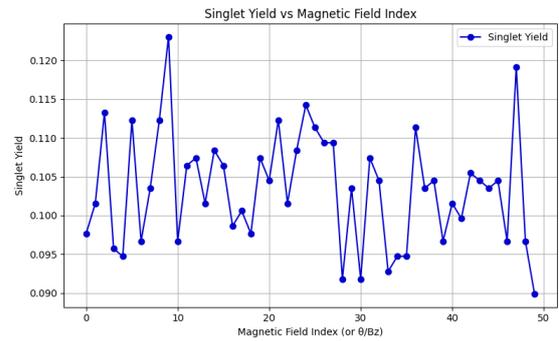

*Figure 11: Singlet yield against magnetic field index*

*Table 2: Qiskit quantum circuit characteristics*

| Items/Quantum Circuits | QC | Transpiled QC |
|---|---|---|
| Depth | 6 | 19 |
| Width | 8 | 131 |
| crx | 2 | |
| Crz | 2 | |
| X | 1 | 2 |
| h | 1 | |
| cx | 1 | |
| z | 1 | |
| Rz | | 23 |
| Sx | | 13 |
| barrier | 1 | 1 |

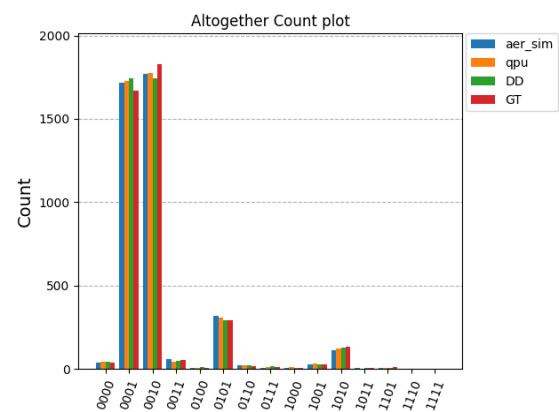

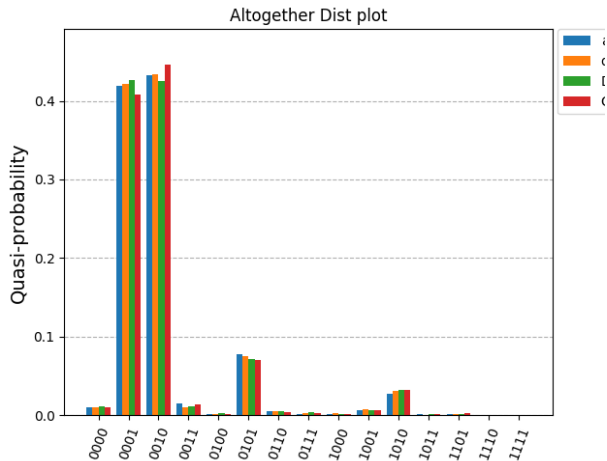

*Figure 13: Distribution plot of different runs in IBM quantum endeavor*

The initial simulation run through IBM Aer Simulator was further extended using the real quantum hardware. Figures 12 & 13 display the comparison histogram and distribution plot of runs in IBM aer simulator, real quantum computer IBM Brsibane, 127 qubits, error mitigation with dynamic decoupling and gate twirling techniques, respectively.

Figure 14 depicts the radical pair singlet-triplet dynamics i.e., probability against a magnetic field range from 10 to 100 μT where the singlet probability dominates above 0.8 whereas triplet probability is negligible in comparison.

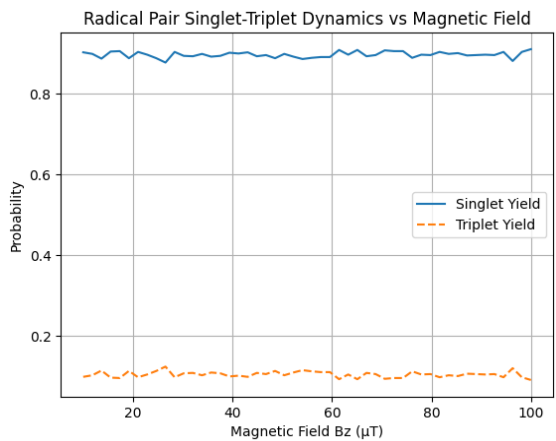

*Figure 14: Radical pair singlet-triplet dynamics over magnetic fields*

Figure 15 reveals the time evolution of the singlet state for all the magnetic field values from 10 to 100 μT plotted at 5 μT intervals. As apparent, the transition almost stabilizes at about 50% probability.

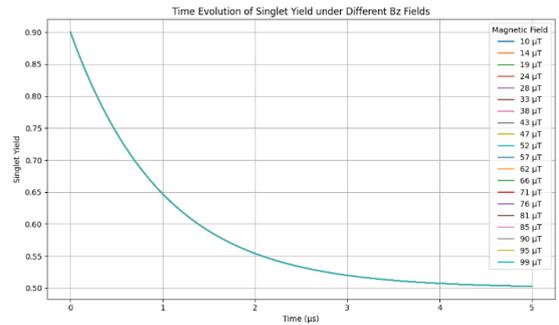

*Figure 15:Time evolution of singlet yields under different magnetic fields*

The purity of electron subsystem found as 0.899 and the state city plot Figure 16, depicting the electron-nuclear correlation which further indicates the radical pair mechanism.

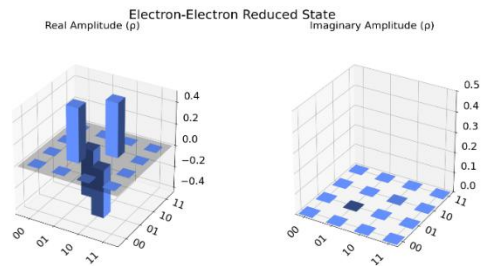

*Figure 16:Tomography of electron-electron sub-system*

Further, a noise model was also built using the Qiskit available libraries. A thermal relaxation error as decay error for qubit relaxation times 50e-03 and 30e-03, respectively, has been introduced for noise modelling purposes. Figures 17, 18, and 19 are the emulating Figures 10, 11 & 14, which resemble each other's trends, ratifying the radical pair-based magnetoreception assumptions.

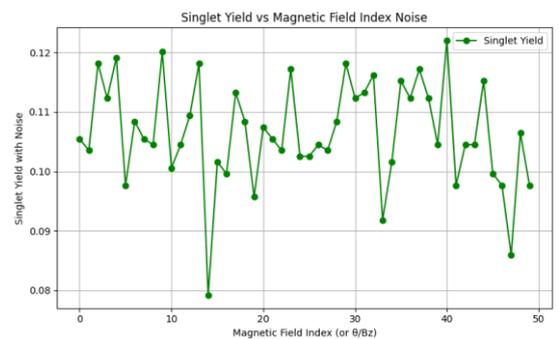

*Figure 17: Noise-model based singlet yield against magnetic field index*

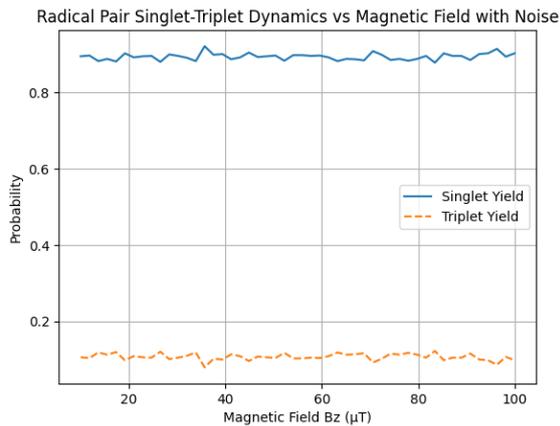

*Figure 18:Noise-model radical pair Singlet-Triplet dynamics*

These Figures provide the noise model simulation identical with the base model run with 1024 shots. Both the set of Figures are identical implying that environmental and other factors does not infringe the characteristics of the radical pair mechanism significantly.

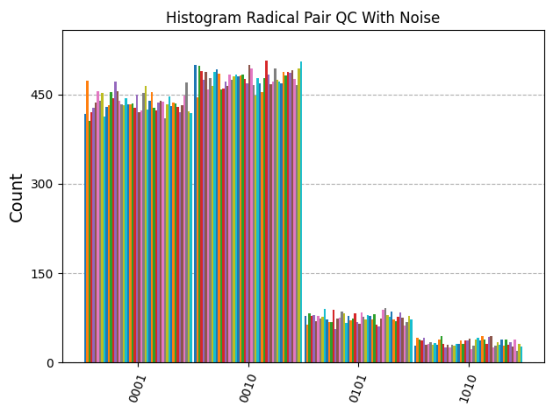

*Figure 19:Noise-model histogram of quantum circuits*

Singlet- Triplet typical transition model with magnetic field gradient in Qiskit reveals the transition with time evolution ideally, Figure 20. It is noteworthy that both the curves meet at mid-point i.e., at about 50% probability.

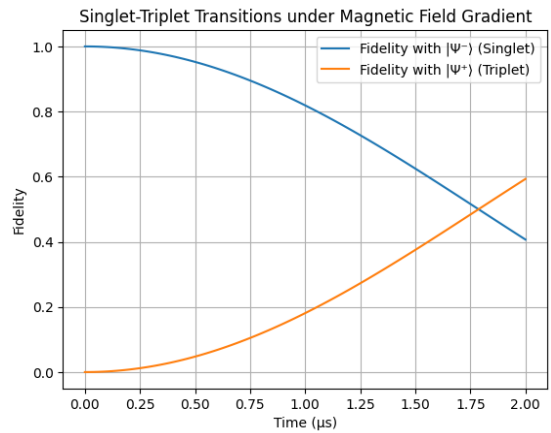

*Figure 20:Ideal singlet-triplet transitions under magnetic field gradient*

The three-dimensional graphs, Figure 21, simulates over a magnetic field range of 0 to 100 µT the transpiring of fidelity, entanglement entropy, mutual information and total correlation. Comparing with the above results, for the same field range singlet-triplet pattern, Figure indicates the coherence apparently maintaining the radical pair mechanism.

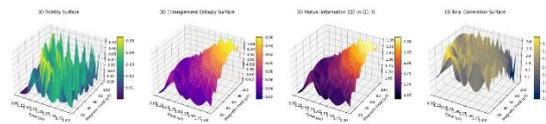

*Figure 21: 3D plots of fidelity, entanglement entropy, mutual information and total correlation*

The study entailed building quantum circuits with four qubits comprising with 2 each for electron-electron and nuclear-nuclear sub-system for a magnetic field range from 10 µT to 100 µT for simulating the earth geomagnetic field. The full system of quantum circuits then run for fifty number of IBM quantum simulator run each with 1024 number of iterations with the base model. Further, considering real-world scenario, a noise model with thermal relaxation error was also built for comparison purpose. The resulting singlet-triplet states, coherence, time evolution appear successful simulation of radical pair mechanism for avian magnetoreception.

**Discussion:**

As evident from the above study, quantum circuit based radical pair mechanism have been modelled under different criteria. For magnetic field strength range from 10 to 100 µT, the 50 number of iterative quantum circuits has been run through IBM Aer Simulator and IBM real quantum hardware. Moreover, error mitigation is also run through dynamic decoupling and gate-twirling methods for comparison. A standalone noise model was also run to compare with the simulator model quickly. Earth magnetic fields, electron gyromagnetic ratio, and Zeeman field splitting have been considered for building each quantum circuit. Figure 13 depicts the probability distribution of four quantum states under different magnetic field strengths, each having distinct probabilities due to different quantum run-through simulators, real quantum computers, and error mitigation techniques. The dominant quantum state appears consistent for all the other magnetic fields. It supports the concept that birds' radical-pair-based magnetoreception is sensitive to minor fluctuations in the geomagnetic field. State probabilities reveal how quantum spin dynamics evolve under varying magnetic fields, mimicking how birds utilize Earth's magnetic field for navigation. With the change in magnetic field strength, probability varies, suggesting that field strength variation affects quantum coherence and spin dynamics simulation. Overall, the results of various runs in different quantum computing techniques are consistent, upholding the idea of radical pair mechanisms in migratory bird magnetoreception.

There has been emerging interest in quantum entanglement in radical pairs in quantum biology. A radical pair's singlet state is entangled, meaning their behavior remains correlated even though the electrons are spatially separated. However, whether entanglement offers a practical benefit in biological magnetoreception is a crucial subject. Moreover, the ability of migratory birds to become confused by weak time-dependent magnetic fields is one of the best arguments in favor of the radical-pair mechanism of magnetoreception. These fields disrupt the birds' visual system's natural processing of geomagnetic fields [12]. In real-world scenarios, migratory birds adapt to numerous factors contributing to their unique generational journey, enabling them to sustain themselves in different parts of the world. The weather and climatic variations are the most prominent factors; however, wind speed, direction, turbulence, the Earth's geomagnetic field, flying formation, clouds, rain, humidity, lightning, atmospheric temperature, and many other natural factors also play a role. Since the Earth's geographic profile is undulated depending on the terrain type, adjusting flight altitude appears to be a real-world scenario that flocks of birds tend to encounter and adapt to during their long journeys. The adaptation might include generational learning of the individual species that comprise millions, if not billions, of them, acquired over ages by different species. Artificial features, such as man-made structures, civilizational or industrial developments, including overhead electricity lines, telecom towers, deep space research facilities that utilize different antennas, ocean developments and explorations, such as offshore wind farms, and many others, are factors that could also plausibly influence the scenario.

The quantum simulation models the radical-pair mechanism, a commonly recognized theory for avian magnetoreception. The findings shed light on the long-term effects of an external magnetic field (B-field) on quantum states as detailed below:

1. Distribution of Quantum States

The bar chart shows the probability distribution of several quantum states (0000, 0100, 1100, and 1000). The modest differences in probability seen at different B-field intensities (10 μT to 100 μT) suggest that the external field influences the evolution of the radical pair.

2. Mechanism of Magnetoreception

An entangled radical-pair state is produced via the Hadamard and CNOT gates. The influence of an external B-field on electron spins is modeled by the Zeeman interaction (Rz rotation). As the minor variations in probability distributions indicate, certain quantum states are more favored under particular B-field values. This could impact the rates of biochemical reactions in biological systems.

3. Physical Interpretation of Sensitivity to Magnetic Fields

Birds are thought to be able to sense changes in the Earth's magnetic field, which is between 25 and 65 μT. The simulation's quantum effects suggest that various magnetic fields influence spin dynamics and quantum coherence, potentially leading to variations in the reaction speeds of biological systems. Due to these effects, birds may be able to navigate by exploiting the Earth's weak magnetic field.

4. Decoherence's Effect (Noise Model)

T1 (relaxation) and T2 (decoherence) are included to simulate actual quantum noise. Over time, thermal relaxation weakens coherence, resulting in a distribution where noise effects make certain states more prevalent. This supports the notion that birds' magnetoreception must occur on brief timescales before decoherence takes over.

## 5. Perspectives on Temporal Development

The probability distribution varies when the system is evolved under different B-fields, illustrating how magnetoreception may be impacted at varying elevations or locations. Therefore, the quantum impact is biologically significant because the system maintains quantum coherence sufficiently long to influence chemical reactions. However, the study of building a quantum circuit envisioned at this stage is for a single bird, irrespective of the differences in species. For multiple birds, especially flocks and formations, the study may include quantum machine learning based on quantum circuits, which is being viewed as a possible future direction of this work. Thus, further developing multiple bird flight formations, based on available data and species-wise, might offer some broader insights, considering magnetoreception and olfaction in combination or parallel, and is also being viewed as a future direction of this work. Moreover, the study might indicate or absent some intriguing thoughts of any role of consciousness, if at all, consciousness would likely have any connection of that of human consciousness, a study gained momentum from hard problems coined by David Chalmers and adjoining neuroscientific endeavors which is a very different aspect might demand into a broader and separate domain intersection. How to feel like a bat, Thomas Nagel [30] raised it, though, from a bat perspective.

### Future Directions

Moreover, many kinds of datasets are available for bird research. The Global Collection of Directional Migration Networks of Migratory Birds consists of 42,844 geo-referenced records that provide subnational-level information on migration times, starting and ending places, and routes [31]. Patterns of Bird Migration in the Western Hemisphere: This dataset, made available by NOAA's Science on a Sphere, provides information on the seasonal trajectories and migratory paths of many bird species in the Western Hemisphere [32]. Species We Study: Birds: The U.S. Geological Survey (USGS) offers migration datasets, tracking flight patterns and behaviors worldwide through GPS transmitters and bird banding [33]. The BirdFlow Project: This NSF-funded partnership between UMass and other organizations models and infers migratory bird population movements using citizen science data, providing information about the birds' migratory paths [34]. Connections: Bird Migration Explorer. Audubon's Bird Migration Explorer offers tracking statistics that show worldwide migratory linkages, produced using technology such as GPS satellite telemetry [35]. The U.S. Fish and Wildlife Service hosts the Migratory Bird Data Center, which provides access to several databases on migrating birds, including point count information and other monitoring initiatives [36]. Plotting flight pathways and determining daily mean speeds during migration are only two of the analyses of bird flight trajectories available in this GitHub repository [37]. Live Bird Migration Maps: The U.S. weather surveillance radar network's nightly bird migration intensities are displayed on BirdCast's real-time analysis maps [38]. The study might need to integrate with either or any of these datasets for generalization purposes.

Using either of these data coupled with quantum computing, further studies should contribute to widening knowledge about the patterns, especially the time duration and distance of flight. Planning of flight according to the flight path i.e., similar to waypoints of aerial commercial flights to address some of the queries for having a realistic albeit generalized understanding of the topic:

Do they always fly in the daytime? What if light-based magnetoreception is the driving reason? Should the night-time flight be considerable?

What is the average aerial speed of flight, albeit it depends on the species?

Is the total flight path continuous over land and sea? If so, what distance of sections typically divides the total path for them to realistically 'plan' the journey?

What is the total journey time in terms of days?

In the follow-up, the available dataset-based studies can plausibly shed more light on this fascinating topic, which has been identified as a future direction to corroborate quantum computing modeling and further extend the studies. As different species tend to travel in flocks and formations, identifying realistic areas and considering further factors in the study can reveal a new dimension in this interesting research domain.

### IV.   CONCLUSION

This quantum computing-based approach addresses an interesting topic of radical pair-based magnetoreception. The unique modeling criteria of superposition, entanglement, and coherence demonstrate the uniqueness of quantum computing application areas, having the potential to spread novel areas. Moreover, using a significant number of quantum circuits and corresponding runs in different quantum techniques, with thousands of shots each, demonstrates the plausibility of Qiskit's use in data streaming situations to address real-world data-driven problems.